\documentclass[final,3p,times,authoryear]{elsarticle}

\usepackage{amssymb}
\usepackage{graphicx}%
\usepackage{multirow}%
\usepackage{amsmath,amssymb,amsfonts}%
\usepackage{amsthm}%
\usepackage{mathrsfs}%
\usepackage[title]{appendix}%
\usepackage{xcolor}%
\usepackage{textcomp}%
\usepackage{manyfoot}%
\usepackage{booktabs}%
\usepackage{algorithm}%
\usepackage{algorithmicx}%
\usepackage{algpseudocode}%
\usepackage{listings}%
\usepackage{tikz}
\usepackage{url}
\usepackage{array} 
\usetikzlibrary{trees, fit, backgrounds, shapes}
\usepackage{subcaption} 

\journal{Knowledge-Based Systems}

\begin{document}

\begin{frontmatter}



\title{LLM-based IR-system for Bank Supervisors}
\date{May 09, 2025}


\author[ecb]{Ilias Aarab\corref{cor1}\fnref{fn1}}
\ead{ilias.aarab@ecb.europa.eu}
\cortext[cor1]{Corresponding author}
\fntext[fn1]{The views expressed in this paper are those of the author and do not necessarily reflect those of the European Central Bank or the Single Supervisory Mechanism.}

\affiliation[ecb]{organization={European Central Bank, SSM/DG-SIB}, addressline={Sonnemannstrasse 20}, 
            city={Frankfurt am Main},
            postcode={60314}, 
            state={Hesse},
            country={Germany}}

\begin{abstract}
Bank supervisors face the complex task of ensuring that new measures are consistently aligned with historical precedents. To address this challenge, we introduce a novel Information Retrieval (IR) System tailored to assist supervisors in drafting both consistent and effective measures. This system ingests findings from on-site investigations. It then retrieves the most relevant historical findings and their associated measures from a comprehensive database, providing a solid basis for supervisors to write well-informed measures for new findings. Utilizing a blend of lexical, semantic, and Capital Requirements Regulation (CRR) fuzzy set matching techniques, the IR system ensures the retrieval of findings that closely align with current cases. The performance of this system, particularly in scenarios with partially labeled data, is validated through a Monte Carlo methodology, showcasing its robustness and accuracy. Enhanced by a Transformer-based Denoising AutoEncoder for fine-tuning, the final model achieves a Mean Average Precision (MAP@100) of 0.83 and a Mean Reciprocal Rank (MRR@100) of 0.92. These scores surpass those of both standalone lexical models such as BM25 and semantic BERT-like models.
\end{abstract}




\end{frontmatter}


\section{Introduction}\label{sec:intro}


Banking supervision in the Euro area is organized in the Single Supervisory Mechanism (SSM), consisting of the European Central Bank (ECB) together with the National Competent Authorities (NCA) of the 20 Euro area countries (as well as Bulgaria through close cooperation). The largest banking groups (and their subsidiaries in the Euro area) are supervised directly by the ECB, while the less significant institutions remain under the supervision of their respective NCAs. Among the responsibilities assigned to it, the ECB as the competent authority, has to approve (material changes to) Supervised Entities’ internal models for the use of own funds requirements calculation. Such approval is usually granted by means of an ECB Decision, being the outcome of an Internal Model Investigation (IMI). The main outcome of an IMI is an in-depth Assessment Report (AR) containing observed findings and their related severity of the inspected model in question. A finding within the AR describes non-compliance with either a legal requirement (e.g., CRR) or a non-binding standard (e.g., EBA Guidelines, ECB Guide to Internal Model), or both. Each finding within the AR is thoroughly and objectively motivated with respect to the violations of the legal frameworks.
This process includes a dedicated and thorough consistency checking phase executed at the ECB.
After the finalization, the AR is shared with the Joint Supervisory Team (JST) that supervises the institution in question. The respective JST takes the AR as main input in order to prepare a Draft Decision (DD) with respect to the approval inquiry to the (material changes of the) internal model. More precisely, the JST links findings within the AR to specific measures, which are further decomposed in conditions, limitations, obligations and recommendations with respect to the (material changes of the) internal model. 
The use of approvals with measures is embedded into the ECBs supervisory approach, and introduces a range of advantages: 
\begin{itemize}
\item Ability to approve a model that does not perfectly meet regulatory expectations, but is better than its predecessor, combined with clearly defined remediation actions
\item Distribution of work between on-site and off-site, i.e., detection of issues via onsite and follow-up of remediation via ongoing model monitoring
\item Ensure adequacy of capital requirements at all times via the use of limitations
\item Ability to compel the Supervised Entity to improve its models via obligations
\end{itemize}
Although clearly of merit, the drafting of the ancillary provision requires a significant effort from the JST in terms of time and needed expertise in order to ensure consistency and objectification of the final results. The goal hereby is to maintain uniformity of IMI decisions within the SSM, i.e., the outcome of these decisions is in line with the treatment given to previous similar cases, as well as to ensure a level playing field across SSM institutions. Auxiliary, the consistency and objectification aids in the prevention of legal and reputation risk for the ECB.\footnote{“This Regulation confers on the ECB… with full regard and duty of care for the unity and integrity of the internal market based on equal treatment of credit institutions with a view to preventing regulatory arbitrage.” Council Regulation (EU) No 1024/2013 Chapter1/Article1}
While being of utmost importance, a consistent conversion of findings into measures is not framed into a strict set of rules given its own very qualitative nature but largely relies on expert judgment and experience of the decision drafter and reviewers. 
In the domain of decision drafting, an important challenge is the write-up of consistent measures conditional on the set of findings found back in the AR of the IMI. One way to ensure such consistency is by comparing a new finding with a set of historical findings and their associated measures. Assessing the relevance of, and the relationship between findings and measures is one of the key skills used by JSTs to ensure consistent writing of Decisions. JST members need to know which previous findings are relevant to a current investigation, in order to draft their Decision and associated measures in a uniform way. More so, they need to assess whether a piece of legislation is relevant to a fact pattern in a finding and determine the effect of the legislation on the measures. Assessing relevance can pose significant challenges, requiring complex analysis and determinations. Automating this task, even partially, could have tremendous implications for JST members across multiple vertical business areas at the SSM.
These implications extend to both the composition of draft decisions and the achievement of a heightened level of consistency. 

In this paper, we present a novel project aimed at augmenting the decision-drafting process through the incorporation of semi-automated tools. This initiative seeks to mitigate existing workload challenges and further bolster the consistency goals. Central to our approach is the development of an Information Retrieval (IR) system, tailored to the specific needs of JSTs. Leveraging Large Language Models (LLMs), this system facilitates efficient retrieval of historical and peer-based findings, alongside their measures, given the new set of findings derived from a new IMI. Our methodology underscores the intersection of applied deep learning techniques with judicial administrative processes, setting a precedent for future innovations in this domain.

The paper is structured as follows. The next section describes related work. Section 3 presents the characteristics of the text data underlying the historical Findings. Section 4 presents the main methodology and experimental setup. Section 5 describes our evaluation methodology, while section 6 summarizes the results of our experiments. Section 7 concludes. 

\section{Related work}\label{sec:related_work}

The field of natural language processing (NLP) and information retrieval (IR) has observed significant advancements in recent years, predominantly due to the introduction of Transformer-based models and novel approaches leveraging deep learning technologies. The seminal work on the Transformer network by \citet{attention} pioneered the transformative shift in NLP. Rejecting recurrence and convolutions, the architecture utilizes self-attention and scaled dot-product attention mechanisms, radically improving efficiency and performance in sequence transduction tasks such as machine translation. The Transformer has been at the core of BERT (Bidirectional Encoder Representations from Transformers) introduced by \citet{devlin2018bert}, which has set profound benchmarks across a spectrum of NLP tasks, including question answering and language inference. BERT innovatively leverages Masked Language Modeling and Next Sentence Prediction, contributing to a deeper contextual understanding of language. Further adaptation of Transformers for specific domains and tasks is exemplified by Legal-BERT \citep{chalkidis2020legal}, which underscores the potential of domain-specific pretraining. Another transformative adaptation can be seen in Sentence-BERT by \citet{reimers2019sentence}, which modifies the base BERT model for efficient computation of sentence embeddings suitable for semantic similarity assessments. 

On the frontier of information retrieval, there is a continuous effort to overcome the limitations of traditional lexical models like BM25. The creation of BEIR, a heterogeneous benchmark by \citet{thakur2021beir}, has enabled a comprehensive zero-shot evaluation of retrieval models across diverse domains, challenging the generalization capabilities of state-of-the-art models. The BM25's capabilities have been expanded in the probabilistic relevance framework presented by \citet{robertson2009probabilistic}. Novel IR frameworks, such as ColBERT by \citet{khattab2020colbert}, leverage a late BERT-based interaction for efficient passage search, balancing between effectiveness and cost-efficiency. This shift towards integrating deep learning within IR systems is further exemplified by UDEG \citep{jeong2021unsupervised}, which employs stochastic text generation to enhance document representation.

Research by \citet{gururangan2020don} has explored the strategic domain-adaptive and task-adaptive pretraining, demonstrating significant progress in the field. Through such adaptations, FinBERT by \citet{araci2019finbert}, specifically pretrained on financial corpora, demonstrates how targeted domain pretraining can yield state-of-the-art results for financial sentiment analysis. Similarly, exploring various strategies to address the challenge posed by very long documents in IR, \citet{lv2011documents} presented BM25L, an extension of BM25, which provided more robust performance across test collections from TREC.

Further contributions include those from \citet{pennington2014glove}, who introduced GloVe, a new global log-bilinear regression model enabling robust word representation learning by capturing global statistics. This advancement along with the CBOW and Skip-gram models by \citet{mikolov2013efficient} represent significant efforts to compute continuous vector representations of words through unsupervised learning.

The unsupervised learning approach has been exemplified by \citet{wang2021tsdae} through the proposal of TSDAE, a Transformer-based sequential denoising auto-encoder that has set new standards in unsupervised sentence embedding learning. This has paved the way for creating semantically rich representations without reliance on labeled data.

The tenets established by foundational works such as those by \citet{sutskever2014sequence} on sequence to sequence learning with neural networks have consequently evolved into advanced methodologies such as Transformers that dominate the current research landscape. These advancements mark significant strides toward a more nuanced and contextually-informed understanding of language, which can be leveraged across a host of applications in both NLP and IR. Based on these advancements we explore the use of Transformer based models to build an IR system for our specific prudential domain. 

\section{Data}\label{data}

\begin{figure}[hp]
 \centering
 \includegraphics[width=.99\textwidth]{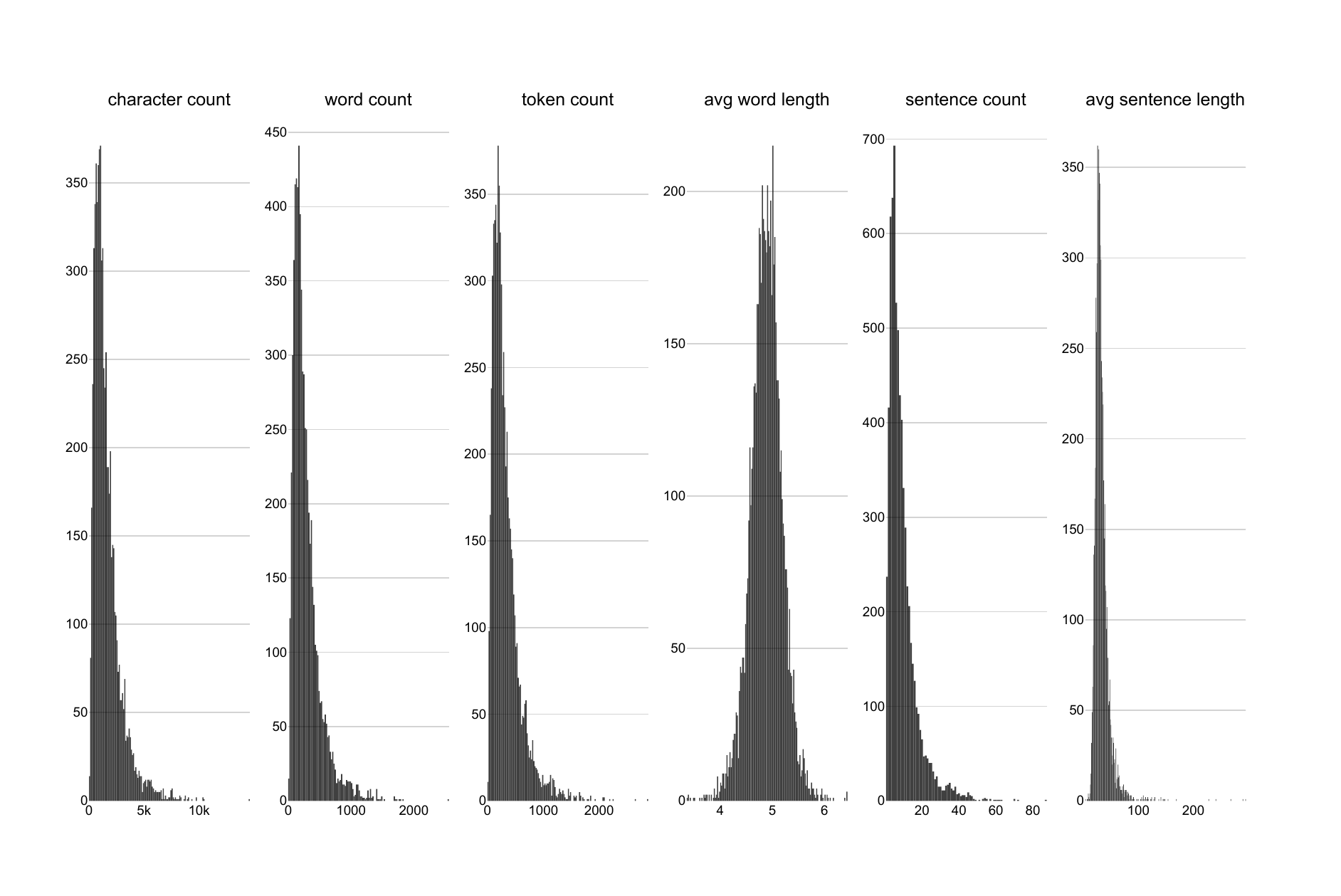}
 \caption{Text statistics of the findings database}
 \label{fig:txt_stats}
\end{figure}

We collect findings from a centralized SSM data lake, focusing specifically on finalized Internal Model Investigations. We exclude data before 2017 to address quality concerns and remove findings linked to problematic IMIs, such as those withdrawn. Our refined dataset consists of approximately 7000 findings. Figure~\ref{fig:txt_stats} shows text statistics of the final database. Notably, around 85\% of these findings are under 512 tokens, fitting the positional embedding size of BERT-based models for straightforward processing. The rest, exceeding this token limit, are segmented based on paragraphs for processing compatibility.

\subsection{Tokenization}

When dealing with Transformer-based models we make use of the accompanying WordPiece tokenizers to process the data before feeding them into the model. These tokenizers are similar to Byte Pair Encoding (BPE) \citep{sennrich2016neural}, but begin the merging process at the word level rather than with individual characters. WordPiece merges the most frequent pairs and uses a greedy algorithm to perform token segmentation in a way that maximizes the likelihood of the training data given the vocabulary \citep{devlin2018bert}.

In contrast, for the lexical models, we employ a custom tokenization scheme. We start with lowercasing and removal of stop words. This step is crucial in eliminating common words such as "the", "and", "is", etc., which occur frequently in the database of findings but often contribute little to understanding the contextual meaning of sentences. For example, consider the stylized sentence:

\

\textit{Institutions shall estimate conversion factors by facility grade or pool on the basis of the average realized conversion factors by facility grade (amidst 2024 planning), pursuant article 182(1)(f) of Regulation (EU) No 575/2013.}

\

The stop words here are "shall", "by", "or", "on", "the", "of", and "by". After this step, the sentence would read:

\

\textit{Institutions estimate conversion factors facility grade pool basis average realized conversion factors facility grade (amidst 2024 planning), pursuant article 182(1)(f) Regulation (EU) No 575/2013.}

\

Following this, we proceed to lemmatize all remaining tokens. This process involves reducing each word to its base or root form, thereby standardizing vocabulary and ensuring consistency in the processing. For example, after lemmatization, we get:

\

\textit{Institution estimate conversion factor facility grade pool basis average realize conversion factor facility grade amidst planning, pursuant article 182(1)(f) Regulation (EU) No 575/2013.}

\

Next, we add bi- and trigrams whenever the conditional probability of two adjacent tokens occurring together is higher than their respective unconditional probabilities\footnote{Probabilities are empirically estimated across the entire Findings database. Trigrams are estimated as the conditional probability of a bigram with an adjacent unigram.}:

\

\textit{Institution estimate conversion\_factor facility\_grade\_pool basis average realize conversion\_factor facility\_grade amidst planning, pursuant article 182(1)(f) Regulation (EU) No 575/2013.}

\

Next we remove tokens that occur in more than 90\% of findings, as these act as in-domain stop words. We also remove tokens occurring in less than .05\% of findings as these are idiosyncratic tokens that do not capture any similarity between findings. In the example this yields the removal of \textit{amidst}, \textit{persuant} and \textit{article}. 

The most distinguishing characteristic of our tokenization scheme is how we handle references to the CRR (Capital Requirements Regulation) articles. We implement a custom regular expression that identifies and correctly tokenizes these references. This ensures that each CRR article reference is treated as a unique token, thereby preserving the specific importance and contextual implications these references carry in financial and regulatory communications.

\section{Methodology \& Experimental Setup} \label{sec:methodology}

Let \(F = \{ f_{0},\ f_{1},\ f_{2},\ \ldots,\ f_{n}\}\) represent the
population of historical findings with the following properties:

\begin{itemize}
\item
  \(|F| = n < \infty\): there is a finite number \(n\) of possible
  findings
\item
  \(|F| \neq \emptyset\): the set is non-empty
\end{itemize}

Let \(f_{n + j}\ \text{for } j = 1,\ \ldots,\ \tau\) represent a new
finding derived from a new IMI, containing \(\tau\) findings.

Furthermore, let \(M = \{ m_{0},\ m_{1},\ m_{2},\ \ldots,\ m_{l}\}\)
represent the population of historical measures with similar properties
as \(F\). Correspondingly,
\(P_{\geq 1}(M) = \{ S:S \subseteq M,\ S \neq \emptyset\}\) represents
the power set of \(M\) excluding the empty set. Then let
\(\theta:F \rightarrow P_{\geq 1}(M)\) represent the set-valued
function, mapping elements from \(F\) onto one or more measures.

Assuming that findings that are similar to each other share similar
measures, a JST member can rely on historical similar findings of
\(f_{n + j}\), and on the observed mapping of \(\theta\), to ensure
a consistent drafting of the new measures based on historical ones.

We implement a simple IR system whose central aspect is the ranking and retrieval of the
top \(k\), with \(k \leq n\), relevant findings. The system processes a user's new finding \(f_{n + j}\) and computes a
similarity score with each finding in the set of historical findings \(F\)
with respect to \(f_{n + j}\). Once each finding \(f_{i} \in F\) has
been scored, the system sorts the findings in descending order based on
their similarity. The top \(k\) findings
\(F_{n + j}^{'} = \{ f_{n + j}^{1},\ f_{n + j}^{2},\ f_{n + j}^{3},\ \ldots,\ f_{n + j}^{k}\}\)
are then retrieved. The user receives this ranked list of findings,
with the system's most relevant results
appearing first.


More formally, the IR system can be
viewed as a general function $\omega$:
\begin{equation}
\omega:F \times F \rightarrow G \subseteq F:{(f}_{n + j},\ F) \rightarrow \omega(f_{n + j},\ F,\sigma,\ k) = F_{n + j}^{'}    
\end{equation}

Typically, the function \(\sigma\) represents some sort of semantic
similarity, i.e., to which extent two documents or passages are
semantically close together. In our work, the similarity between any two
findings reflects the extent to which they elaborate on the same
non-compliance issue(s) as reflected by the Capital Requirement
Regulation (derived from one or more regulatory articles).

Cosine similarity is a popular metric in information retrieval systems, valued for its ability to detect semantic similarities within high-dimensional vector spaces. It calculates the cosine of the angle between two vectors to produce a score that indicates the degree of alignment between the vectors, without being affected by their sizes. This feature is particularly beneficial in text retrieval systems, where document embeddings can vary widely, and the focus is on the document's content direction within the vector space. Research has consistently shown that cosine similarity surpasses other metrics in retrieval efficiency in such contexts  \citep{manning2008introduction}.

Thus, we now have,

\begin{equation}
\sigma:F \times F \rightarrow \lbrack -1,1\rbrack:\ \sigma\left( f_{n + j},\ f_{i} \right): = y_{n + j,i} = \frac{{v}_{n + j}\  \cdot {v}_{i}\ \ }{\left\| {v}_{n + j}\ \  \right\|\left\| {v}_{i}\ \  \right\|}\ \  
\end{equation}

with $v_{n+j}$ and $v_i$ respectively the vector embeddings of findings $f_{n+j}$ and $f_i$, and $y_{n + j,i}$ the similarity score between them. In practice, for computational efficiency, we compute the cosine
similarity of all findings simultaneously by constructing the following
matrices:

\begin{itemize}
\item
  \(\Pi \in \mathbb{R}^{n \times d}\), where each row \(\pi_{i}\)
  represents the normalized embedding vector of the historical
  finding \(f_{i}\) stored in our database.
\item
  \(\Pi' \in \mathbb{R}^{\tau \times d}\), where each row
  \(\pi'_{j}\) represents the normalized embedding vector of
  the set of new findings \(f_{n + j}\).
\end{itemize}

We can then easily compute the similarity matrix:

\begin{equation}
\Sigma = \Pi'\left( \Pi \right)^{T} \in \mathbb{R}^{\tau \times n}
\label{sim_matrix}
\end{equation}

With \(\Sigma_{n + j,i}\) the cosine similarity between a new finding
\(f_{n+j}\) and the historical finding \(f_{i}\),
\(\forall i \in \left\{ 1,\ 2,\ \ldots,\ n \right\},\ \forall j \in \left\{1,2,\ldots,\tau\right\}\).

Building a proficient IR system is a challenging task, primarily due to the necessity of establishing an effective embedding scheme capable of converting findings into meaningful and representative
vectors. We group the embedding schemes into three broad categories: (i) lexical, (ii) word-level embeddings and (iii) document-level embeddings.

\begin{figure}[hp]
 \centering
 \includegraphics[width=.99\textwidth]{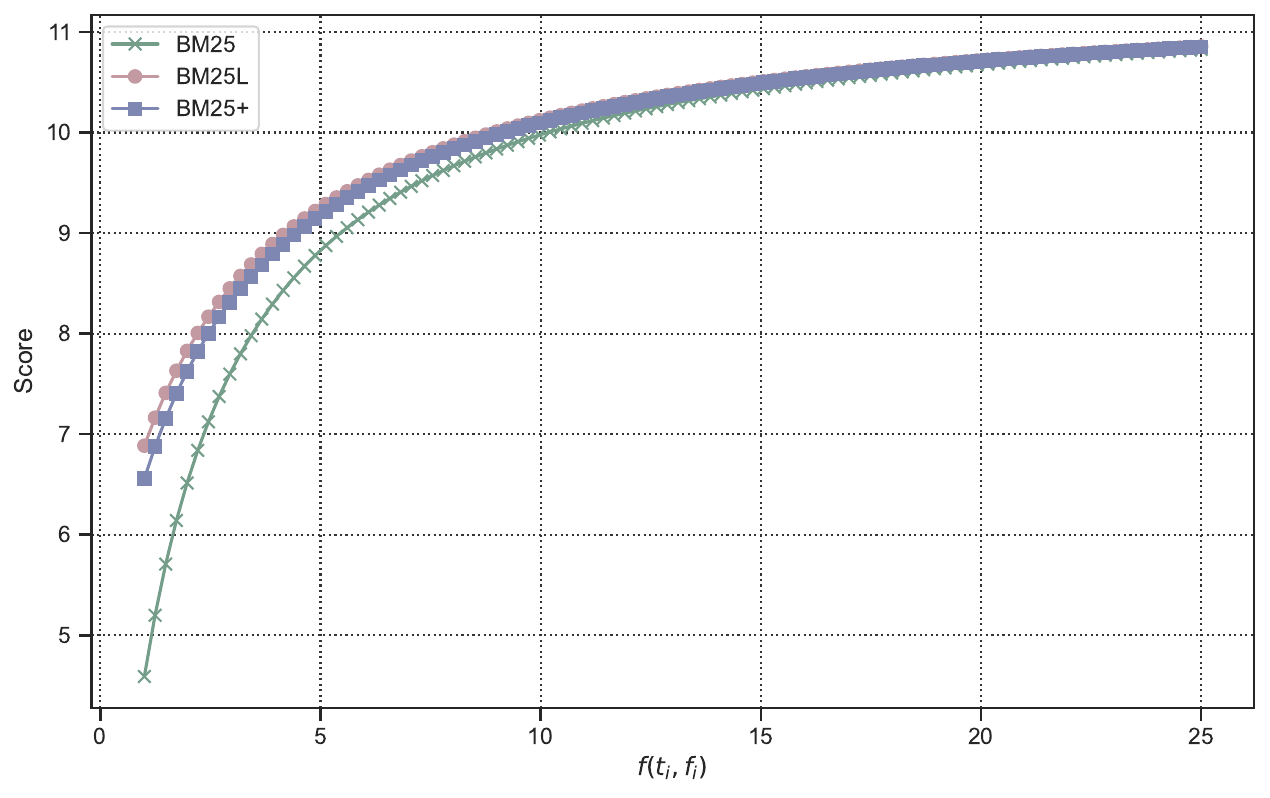}
 \caption{Comparison of BM25, BM25L and BM25+}
 \label{fig:bm25}
\end{figure}

\textbf{Lexical:} The TF-IDF model computes the importance of a term within a finding $f_i$ relative to the entire findings database $F$ , leveraging the term frequency (TF) and the inverse document frequency (IDF) as follows:
\begin{equation}
\text{TF-IDF}(t, f_i, F) = TF(t, f_i) \times IDF(t, F)
\end{equation}
where \(t\) represents the term, \(f_i\) is the finding, and \(F\) is the findings database. TF-IDF increases with the number of occurrences of the term in the finding but is offset by the term's frequency in the finding database, ensuring that common terms are appropriately weighted \citep{SPARCKJONES1972}. BM25 extends upon the basic principles of TF-IDF by incorporating document length normalization and a saturation function, making it less sensitive to term frequency increases beyond a certain threshold. It is defined as:
\begin{equation}
\text{BM25}(t, f_i, F) = \sum_{i=1}^{n} IDF(t_i, F) \cdot \frac{f(t_i, f_i) \cdot (k_1 + 1)}{f(t_i, f_i) + k_1 \cdot (1 - b + b \cdot \frac{|f_i|}{\text{avgdl}})}
\end{equation}
where \(f(t_i, f_i)\) is \(t_i\)'s term frequency in finding \(f_i\), \(|f_i|\) is the length of the document, \(\text{avgdl}\) is the average finding length in the database, and \(k_1\) and \(b\) are free parameters \citep{Robertson2009}. BM25L further refines BM25 by addressing its limitations related to term frequency saturation. It introduces a term frequency normalization step that is less prone to saturation, making it more effective in environments where term distribution is sparse. Its formulation can be viewed as an extension to the BM25 formula with additional normalization factors \citep{lv2011documents}:
\begin{equation}
\text{BM25L}(t, f_i, F) = \sum_{i=1}^{n} \text{IDF}(t_i, F) \cdot \frac{(k_1+1)(c(t,f_i)+\delta)}{k_1 + (c(t,f_i) + \delta)}
\end{equation}
with
\begin{equation}
c(t,f_i) = \frac{f(t_i, f_i)}{1 - b + b \cdot \frac{|f_i|}{\text{avgdl}}}
\end{equation}
and $\delta > 0$ a shift parameter.
In contrast, BM25+ refines the BM25 model by adding a small constant \( \delta \) to the term frequency component, enhancing sensitivity to term frequency without compromising normalization, thereby improving document relevance assessment \citep{lv2011lower}. 
\begin{equation}
\text{BM25+}(t, f_i, F) = \sum_{i=1}^{n} IDF(t_i, F) \cdot (\frac{f(t_i, f_i) \cdot (k_1 + 1)}{f(t_i, f_i) + k_1 \cdot (1 - b + b \cdot \frac{|f_i|}{\text{avgdl}})} + \delta)
\end{equation}
This adjustment allows for more linear scaling with term frequency, addressing term frequency saturation more effectively. Figure~\ref{fig:bm25} highlights the difference between the three BM25 implementations by computing scores of a new finding containing one unique term at different frequencies. As can be seen, both BM25L and BM25+ show a higher tolerance for shorter documents, which results in a generally higher score for the same term frequencies compared to BM25. This makes them particularly useful for environments where document length varies greatly (see Figure~\ref{fig:txt_stats}) but shorter documents still contain valuable content.

\textbf{Word-level Embeddings:} GloVe (Global Vectors for Word Representation) is a model designed to efficiently learn word vectors by aggregating global word-word co-occurrence statistics from a corpus, capturing both local and global semantic relationships between words \citep{pennington2014glove}. Building on this, we propose an extension to adapt GloVe embeddings for the prudential domain by training on a corpus comprising financial reports, regulatory documents, and related texts, thereby embedding financial jargon and context to enhance the relevance of representations for improved information retrieval performance. We further enhance semantic understanding of findings with BERT's contextual embeddings, which unlike GloVe's static representations, dynamically adjust word meanings based on surrounding text, offering nuanced insights \citep{devlin2018bert}. FinBERT \citep{finbert} and LEGAL-BERT \citep{chalkidis2020legalbert} extend this with finance and legal domain specializations, embedding sector-specific contexts for more nuanced document representations. We aggregate GloVe embeddings, by computing the average vector across all word embeddings. For BERT embeddings, we do the same but exclude the [CLS] token embedding.

\begin{figure}[hp]
 \centering
 \includegraphics[width=.99\textwidth]{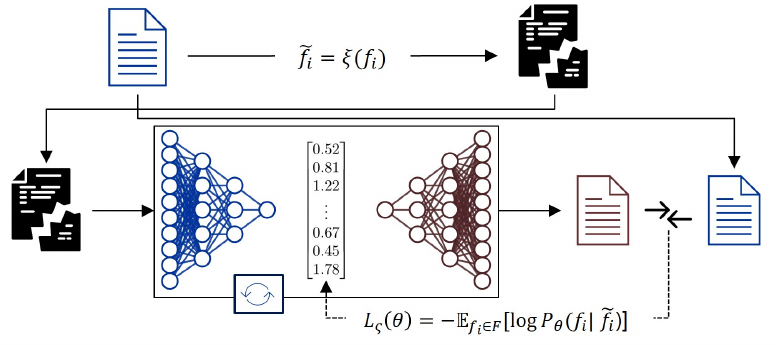}
\caption{The diagram illustrates a transformer-based sequential denoising autoencoder. The process begins in the \textbf{top part} where the input finding $f_i$ is corrupted using the function $\tilde{f_i} = \xi(f_i)$. The corrupted finding $\tilde{f_i}$ is then processed through the \textbf{left middle part}, which contains the encoder network that generates a latent representation. This latent representation (shown in the \textbf{center}) is passed to the decoder network in the \textbf{right middle part} to reconstruct the original finding. The model is optimized by minimizing the reconstruction loss $L_{\zeta}(\theta) = -\mathbb{E}_{f_i \in F}[\log P_{\theta}(f_i | \tilde{f_i})]$, calculated at the \textbf{bottom part} of the diagram, ensuring that the output closely matches the original input after denoising.}
 \label{fig:tsdae}
\end{figure}

\textbf{Document-level Embeddings:} Sentence Transformers \citep{reimers2019sentencebert} advance BERT-like models by generating document-level embeddings, contrasting traditional word-level embeddings by capturing broader semantic contexts. Unlike word embeddings, which focus on individual word meanings, Sentence Transformers embed entire documents, leading to better performance in tasks requiring nuanced comprehension of text, such as document retrieval and question answering. To fine-tune Sentence Transformers, we use a Transformer-based Sequential Denoising Auto-Encoder (TSDAE) \citep{feng2021transformer}. TSDAE trains by reconstructing documents from their noise-altered forms, improving the model's grasp on language structure and semantics without needing labeled data. More formally, let $\varsigma(\cdot)$ represent a pretrained Sentence Transformer. Let $f_i$ be the original finding, and $\Tilde{f}_i=\xi(f_i)$ be a noisy version obtained by applying a noise function $\xi$ that randomly deletes 50\% of the tokens. In the denoising step, we use $\varsigma(\Tilde{f}_i)$ to reconstruct the original tokenization of the finding $f_i$. To learn $\varsigma(\cdot)$ to reconstruct the original finding, we utilize the following loss objective:
\[ L_\varsigma(\theta) = -\mathbb{E}_{f_i\in F}[\log P_\theta(f_i | \Tilde{f}_i)] \]
where $\theta$ represents the set of weights of the Sentence Transformer $\varsigma(\cdot)$, $\mathbb{E}_{f_i\in F}[\cdot]$ denotes the expectation taken over the set of findings, and $P_\theta(f_i | \Tilde{f}_i)$ is the conditional probability under $\theta$ of generating $f_i$ given $\Tilde{f}_i$. The loss function is minimized using backpropagation \citep{attention}. Once trained, the transformer $\varsigma(\cdot)$ can generate embeddings for findings $f_i$ that are specifically adapted to our domain, even in the absence of labeled data (see Figure~\ref{fig:tsdae}). 

The pretrained Sentence Transformer used is all-MiniLM-L6-v2 \citep{reimers2019sentencebert}, a smaller and faster version of BERT designed for sentence embeddings. The all-MiniLM-L6-v2 model consists of 6 layers (transformer blocks) with 8 attention heads per layer, totaling 48 attention heads. To enhance the understanding of $\varsigma(\cdot)$ in the context of prudential regulation, we initially train the network on all CRR articles $C_i$ before incorporating our historical findings. The model is trained using TSDAE for 125 epochs, which corresponds to the point at which our loss reaches its lowest value, with a batch size of 64. 

\subsection{Fuzzy CRR article matching}

Semantic similarity of the findings is only part of the story. Each finding in the database is also linked to a set of Capital Requirements Regulation (CRR) articles\footnote{The Capital Requirements Regulation (CRR) is a part of the European Union's regulatory framework for financial institutions, aimed at ensuring their resilience and stability by setting out prudential requirements for banks and other financial institutions.}. We leverage this metadata to consider the overlap in CRR articles between findings. The more CRR articles two findings share, the higher their Jaccard similarity, indicating that these findings are likely more related to each other. This aspect lends an additional layer of robustness to the information retrieval system.

The Jaccard similarity between two findings is computed based on the CRR articles linked to each finding. Let A and B denote the set of CRR articles associated with two different findings. The Jaccard similarity is given by the size of the intersection of A and B divided by the size of their union:

\[ J(A,B) = \frac{{|A \cap B|}}{{|A \cup B|}} \]

Similar to cosine similarity, this measure ranges from 0 to 1. A value of 0 implies no shared articles between the findings, while a value of 1 implies that all articles linked to one finding are also linked to the other.

The Jaccard similarity is particularly advantageous as it considers both the similar and different articles between two findings, unlike the Szymkiewicz-Simpson coefficient, which only highlights the overlapping set between two findings. This distinction is important because the different articles can indicate differences in findings. Similarly, the Sørensen-Dice coefficient may dilute the similarity between findings, especially when some findings have a large number of associated articles, as it focuses on the full cardinality of each set of articles. In contrast, the Jaccard similarity considers the union of the sets, making it less susceptible to this issue. 

We further model the hierarchical structure of CRR articles as a directed rooted tree. Each CRR article is a node in the tree, with edges representing the parent-child relationships between articles. For instance, articles 181(a) and 181(b) are child nodes of the parent node 181. The root of the tree represents the highest level of the CRR hierarchy.

Given this structure, we can define a similarity measure based on shared ancestry. Let P(x) denote the set of parent nodes (ancestors) of a node x in the tree. The hierarchical similarity between two nodes x and y can then be defined as:

\[ H(x,y) = \frac{{|P(x) \cap P(y)|}}{{|P(x) \cup P(y)|}} \]

This measure, similar to the Jaccard similarity, ranges from 0 (no shared ancestors) to 1 (all ancestors are shared), providing a quantification of the degree to which two findings are related through their associated CRR articles. By considering both Jaccard and hierarchical similarities, we can achieve a more nuanced understanding of the relationship between findings based on their associated CRR articles. 

\subsection{Experimental setup}

Consider a test dataset $\mathcal{D}$ comprising labeled findings, intended for evaluating the efficacy of embedding schemes. To establish a benchmark for comparison, we introduce a naive IR system that randomly retrieves a set of findings from $F$, based on a slight modification of the hyper geometric distribution as depicted in \citep{Bestgen2015}. Furthermore, we introduce a hybrid model based on the average of similarity matrices $\Sigma$ (see Equation~\ref{sim_matrix}) , derived from both a fully configured \textit{BM25L+} model and a fine-tuned sentence transformer model \textit{SentTRF+TSDAE}, dubbed \textit{Hybrid}. The fuzzy CRR matching component reduces the search space by only including findings \( f_i \) that meet specific criteria: a CRR Jaccard similarity of \( J(A,B) \geq \frac{1}{3} \) and a hierarchical similarity of \( H(A,B) \geq \frac{1}{3} \). When the IR system processes a finding \( f_i \), it first limits the search to findings that satisfy these criteria. After narrowing down the search space, the system then proceeds to identify similar findings within this constrained set. Table ~\ref{tab:experiment_setup} gives an overview of the different embeddings used.

\begin{table}[ht]
    \centering
    \begin{tabular}{l|p{10cm}}
    \textbf{Embedding Scheme} & \textbf{Details} \\
    \hline
    \textbf{Lexical} & \\
    TFIDF & Default implementation. \\
    BM25 & Parameters: $k=1.6$, $b=0.75$. \\
    BM25Plus & Parameters: $k=1.6$, $b=0.75$, $\delta=1$. \\
    BM25L & Parameters: $k=1.6$, $b=0.75$, $\delta=0.5$. \\
    BM25L+ & Parameters: $k=1.6$, $b=0.75$, $min\_df=0.0005$, $max\_df=0.9$, $ngram=3$. \\
    \textbf{Word-level Embeddings} & \\
    GloVe & Dimension: $d=300$, Pooling: Mean. \\
    FinGloVe & Dimension: $d=300$, Pooling: Mean. \\
    BERT & Configuration: bert-base-uncased, Dimension: $d=512$, Pooling: Mean. \\
    FinBERT & Configuration: finbert, Dimension: $d=512$, Pooling: Mean. \\
    LEGAL-BERT & Configuration: legal-bert-base-uncased, Dimension: $d=512$, Pooling: Mean. \\
    \textbf{Document-level Embeddings} & \\
    SentTRF & Model: all-MiniLM-L6-v2, Dimension: $d=384$. \\
    SentTRF+TSDAE & Model: all-MiniLM-L6-v2, Dimension: $d=384$, Training Corpus: $F$ + $CRR$ \\
    Hybrid & Hybrid Model: $\Sigma = \frac{(\Sigma_{BM25L+} + \Sigma_{SentTRF+TSDAE})}{2}$.
    \end{tabular}
    \caption{Overview of Evaluated Embedding Schemes. \textit{min\_df} and \textit{max\_df} represent respectively the minimum and maximum document frequency that a token can have. Pretrained Transformer models are pulled from HuggingFace \citep{HF}. The Pretrained GloVe model is downloaded from Stanford's NLP website (\protect\url{https://nlp.stanford.edu/projects/glove/})}
    \label{tab:experiment_setup}
\end{table}

\section{Evaluation methodology}\label{eval}

\subsection{Evaluation metrics}

Evaluating the performance of an IR system necessitates utilizing a variety of metrics to gain a comprehensive understanding of its effectiveness. In our research, we primarily focus on two metrics: Mean Average Precision (MAP) and Mean Reciprocal Rank (MRR). Both MAP and MRR are extensively recognized and employed in the assessment of IR systems. MAP provides a measure of precision across recall levels, while MRR offers insight into the rank position of the first relevant item in the retrieved results. These metrics are critical in understanding the overall performance of IR systems, as demonstrated in \cite{shah2004evaluating}.

MAP is a standard measure for IR tasks, particularly important in scenarios where the order of returned documents is crucial. MAP calculates the mean of average precisions across all queries. The average precision of a query is computed as the average of precisions at the positions of each relevant document within the ranked retrieval results. By aggregating precision scores at each relevant document, MAP effectively encapsulates both precision and recall, providing a comprehensive single-figure measure of the quality of ranked retrieval outcomes. More precisely, for each new finding \(f_{(n+j)}\), the average precision (AP) is given by:
\[
AP = \int_0^1 p(r) dr
\]
Here, AP represents the area under the precision-recall curve for every position in the ranked sequence of returned historical findings by the IR system. The MAP score is then the mean of the APs over all findings \(f_{(n+j)}\) for \(j \in 1, \ldots, \tau\):
\[
MAP = \frac{{\sum_{j=1}^\tau AP(f_{(n+j)})}}{\tau}
\]
The MAP score is generally calculated up to a certain threshold \(k\) of the number of retrieved documents, often expressed as MAP@k \cite{joshi2012stop}.

On the other hand, MRR is a statistical measure for evaluating any process that produces a list of possible responses to a sample of queries, ordered by probability of correctness. The reciprocal rank of a query response is the multiplicative inverse of the rank of the first correct answer. MRR is the average of these values across queries \cite{shah2004evaluating}. This metric is particularly useful when the most relevant documents are desired at the top of the ranking list, which is a desired property of the IR system we develop. Mathematically, the reciprocal rank (RR) for a single finding \(f_{(n+j)}\) is simply \(\frac{1}{{\text{{rank}}_{(f_{(n+j)})}}}\), with \(\text{{rank}}_{(f_{(n+j)})}\) defined as the position of the first retrieved similar finding by the IR system. MRR is then simply:
\[
MRR(f_{(n+j)}) = \frac{{\sum_{j=1}^\tau RR(f_{(n+j)})}}{\tau}
\]

Both MAP and MRR have been used extensively in the field of information retrieval, allowing for meaningful comparisons with other systems and benchmarks. Their widespread acceptance in the research community lends credibility and comparability to our evaluation process.

\subsection{Validation approach}

In the context of IR systems, it is customary for researchers to utilize standardized labeled data from sources like TREC (Text Retrieval Conference) for model validation \cite{voorhees2005trec}. These datasets are manually labeled by experts, allowing for the evaluation of systems across a diverse set of queries and documents. However, their coverage of specific domains, such as the prudential domain in our study, may be limited \cite{lupu2009trec, abacha2019question}. Given our explicit goal to assess the performance of our system on prudential data, relying on these generic datasets might lead to biased or unrepresentative results. Thus, we need to devise a methodology to work with our unlabeled data, a situation frequently encountered in domain-specific applications due to the prohibitive costs of exhaustive manual labeling \cite{zhou2007knowledge}.

One of the significant challenges in creating a labeled test dataset for our IR system is the need to label all findings in the database as either relevant or non-relevant compared to a test finding, and this for all test findings. This task is resource-intensive and can be prohibitively expensive \cite{koh2009test, duh2008learning}. A straightforward solution is to label only a portion of the database as relevant compared to a test finding, however, this approach introduces a risk of not identifying some genuinely relevant findings. As a result, an ideal IR system that retrieves these non-identified relevant findings would be erroneously penalized, creating an illusion of the system ranking irrelevant findings high. To address these issues, we propose a simple validation approach that accommodates the uncertainty of non-identified relevant findings.

\

Let $f_j$ be a test finding. Let $G=\{f_i \in F | \sigma(f_i, f_j) \geq T\}$ be the true set of similar findings relative to $f_j$, with $T \in [0,1]$ a theoretical threshold value indicating that two findings are similar to each other. Then, let $\hat{G} \subseteq G$ be the set of identified similar findings, so that $\tilde{G}=G \setminus \hat{G}$ is the remaining set of non-identified similar findings related to $f_j$. Lastly, let $D=\{f_i \in F | f_i \notin \hat{G}\}$ be the set of historical findings excluding those findings that are identified as similar to $f_j$. 

Assume the existence of a perfect IR system. In case we test the performance of the system against $f_j$, the first set of retrieved findings would be $G$, whereas the labelled dataset would only classify the subset $\hat{G}$ as correctly retrieved findings and would wrongly punish the system for retrieving the subset $\tilde{G}$.

Now, assume $|D \setminus \tilde{G}| \gg |\tilde{G}|$, such that $p(\tilde{G}|D) \approx 0$, i.e., if the number of historical findings that are truly not similar to the test finding $f_j$ is much larger than the number of non-identified similar findings, then the conditional probability of randomly drawing from $D$ a sample that is also a non-identified similar finding becomes very low. More formally: Let $Q = \{f_i \in D| f_i \notin \tilde{G}\}$ for ease of notation. 
If we keep $|\tilde{G}|$ fixed and let $|D|$ grow to infinity, then the conditional probability $p(\tilde{G}|D)$ goes to zero, as
$$
\lim_{(|D| \to \infty)} p(\tilde{G}|D) = \lim_{(|Q| \to \infty)} \frac{|\tilde{G}|}{|\tilde{G}| + |Q|} = 0,
$$ noting that $D=\tilde{G} \cup Q$ with $Q \cap \tilde{G} = \emptyset$.

Similarly, if we keep $|Q|$ fixed and let $|\tilde{G}|$ shrink to zero, then the conditional probability $p(\tilde{G}|D)$ goes to zero, as
$$
\lim_{(|\tilde{G}| \to 0)} p(\tilde{G}|D) = \lim_{(|\tilde{G}| \to 0)} \frac{|\tilde{G}|}{\tilde{G}| + |Q|} = 0.
$$

This reflects the intuitive idea that the less of $D$ is made up of $\tilde{G}$, the less likely we are to draw an element from $\tilde{G}$. This happens both when $\tilde{G}$ gets smaller and when $Q$ gets larger, assuming that we're drawing uniformly at random from $D$.

Based on the above, we introduce the following simple Monte Carlo sampling method \cite{hastings1970monte, burgin1999monte}:

\ 

\textbf{Step 1:} Let $d_i=\{d_i^1, d_i^2, d_i^3, \dots, d_i^m\} \sim U(D)$ be a random uniformly distributed subset of $D$ of size $m$.

\textbf{Step 2:} We then add the identified similar findings to the random subset, such that $\bar{d}_i = d_i \cup \hat{G}$. where $\bar{d}_i$ can be seen as a down-sampled version of the full historical database of findings.

\textbf{Step 3:} We repeat this process $M$ times, creating $M$ different down-sampled versions of the historical database.

\textbf{Step 4:} For each of the down-sampled versions of the database, we test the performance of the system in retrieving similar findings of $f_j$ by computing $MAP$ and $MRR$ scores.

\textbf{Step 5:} We average the $MAP$ and $MRR$ values across the $M$ down-sampled sets to obtain the final performance figures for $f_j$.  

\

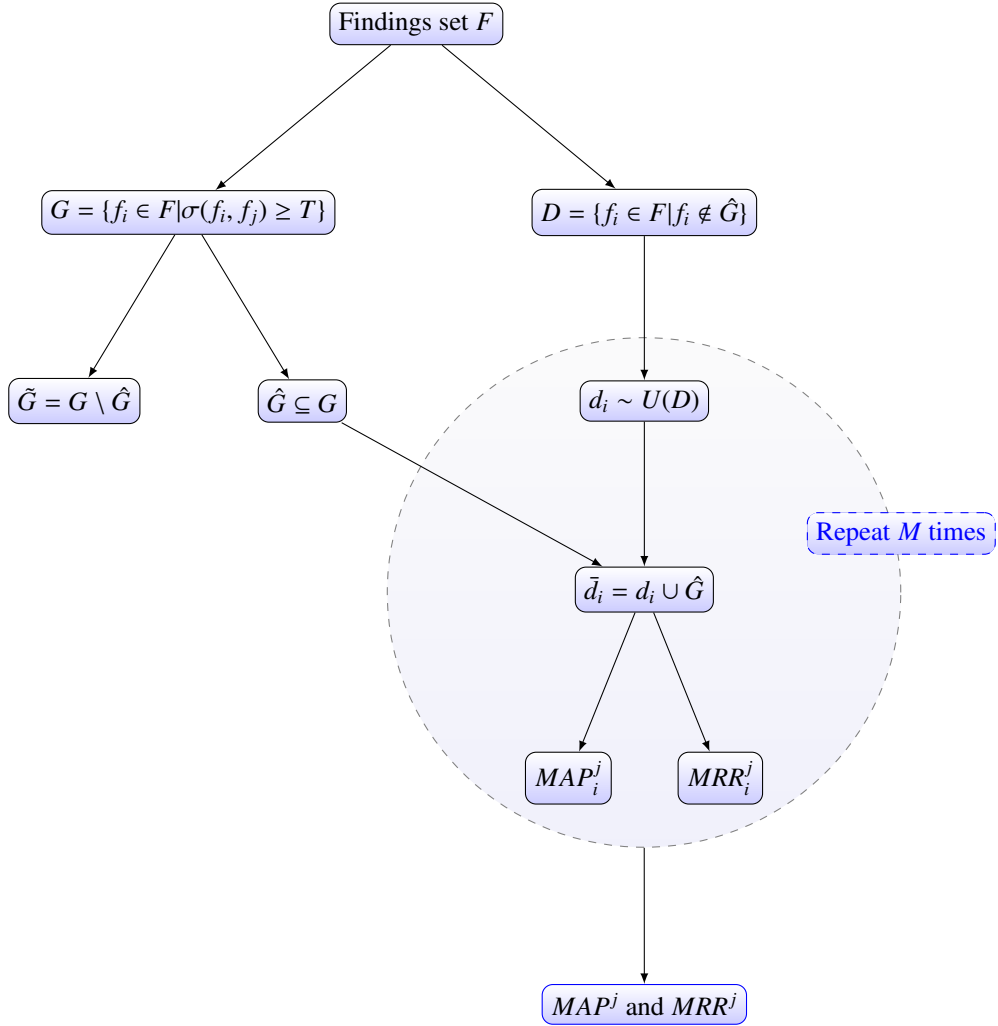
\begin{figure}[h]
\centering
\begin{tikzpicture}[level distance=2.5cm,
  level 1/.style={sibling distance=6cm},
  level 2/.style={sibling distance=3cm},
  level 3/.style={sibling distance=2cm},
  every node/.style = {shape=rectangle, rounded corners,
    draw, align=center, top color=white, bottom color=blue!20},
  edge from parent/.style={draw, -latex},
  annotation/.style = {shape=rectangle, draw=black, dashed, align=center, fill=green!20, color=blue!100},
  finalscore/.style = {rectangle, draw=blue, align=center, fill=blue!20}]

  \node {Findings set $F$}
    child { node {$G = \{f_i \in F | \sigma(f_i, f_j) \geq T\}$}
      child { node {$\tilde{G} = G \setminus \hat{G}$} }
      child { node (Ghat) {$\hat{G} \subseteq G$} } }
    child { node {$D = \{f_i \in F | f_i \notin \hat{G}\}$}
      child { node (Di) {$d_i \sim U(D)$}
        child { node (Dbardi) {$\bar{d}_i = d_i \cup \hat{G}$}
          child { node (MAPIJ) {$MAP_i^j$} }
          child { node (MRRIJ) {$MRR_i^j$} }
        } 
      } 
    };

  \draw[-latex] (Ghat) -- (Dbardi);

  \begin{scope}[on background layer]
    \node (circle) [circle, draw=black!50, dashed, minimum size=4.5cm, fill=black!10, fill opacity=0.2, fit=(Di) (Dbardi) (MAPIJ) (MRRIJ)] {};
  \end{scope}

  \node at (circle.east) [annotation, above=0.5cm] {Repeat $M$ times};

  \node at (3,-13) [finalscore] (MAPj) {$MAP^j$ and $MRR^j$};

  \draw[-latex] (circle) -- (MAPj);

\end{tikzpicture}
\caption{Birdview of the validation methodology.}
\label{fig:val_meth}
\end{figure}

Now we can safely test the IR system by averaging across the down-sampled versions $\bar{d}_i$ instead of the full database $F$. This approach should offer a better estimate of its performance, as the down-sampled versions would contain only those relevant findings that we have actually identified as relevant in case $D$ was truly infinitely large.
While both $\tilde{G}$ and $D$ are finite in practical scenarios, the approximation method still holds provided the cardinality of $\tilde{G}$ is much less than $D \setminus \tilde{G}$, i.e., $|\tilde{G}| \ll |D \setminus \tilde{G}|$. Figure~\ref{fig:val_meth} shows a birdview of the methodology, and Table~\ref{tab:sets_summary} shows an overview of the different set definitions used.

\begin{table}[h]
\centering
\resizebox{\textwidth}{!}{%
\begin{tabular}{|c|>{\centering\arraybackslash}m{5cm}|>{\centering\arraybackslash}m{5cm}|}
\hline
\textbf{Set Symbol} & \textbf{Definition} & \textbf{Description} \\
\hline
$F$ & Findings set & The complete set of findings \\
\hline
$G$ & $\{f_i \in F \mid \sigma(f_i, f_j) \geq T\}$ & Set of true similar findings relative to $f_j$ \\
\hline
$\hat{G}$ & $\hat{G} \subseteq G$ & Set of identified similar findings \\
\hline
$\tilde{G}$ & $G \setminus \hat{G}$ & Set of non-identified similar findings \\
\hline
$D$ & $\{f_i \in F \mid f_i \notin \hat{G}\}$ & Set of historical findings excluding those in $\hat{G}$ \\
\hline
$Q$ & $\{f_i \in D \mid f_i \notin \tilde{G}\}$ & Theoretical ground set of non-similar findings relative to $f_j$ \\
\hline
$d_i$ & $d_i \sim U(D)$ & A random uniformly distributed subset of $D$ of size $m$ \\
\hline
$\bar{d}_i$ & $d_i \cup \hat{G}$ & Down-sampled version of the full historical database \\
\hline
\end{tabular}
}
\caption{Description of the sets utilized within the validation methodology.}
\label{tab:sets_summary}
\end{table}

In order to further quantify the deviation from the theoretical setup, we conduct the following simulation exercise:
Firstly, an artificial database of documents ($db$) is created, having the same size as our findings database $F$, such that $|db|\approx7000$. An artificial test finding $t$ is also introduced. 

The constructed database $db$ and test finding $t$ possess certain properties: the sets $G$ and its subsets $\hat{G}$ and $\tilde{G}$ are known and adjustable in terms of their sizes. This is unlike our real database $F$, where $\tilde{G}$ is unobservable. Therefore, both $D$ and $D \setminus \tilde{G}$ are known and their sizes can be adjusted.
We then introduce a perfect IR system, $\omega_1(.)$, which greedily retrieves all similar findings of $t$ such that, $\omega_1(t,db)=\{g_1^{\tilde{G}}, g_2^{\tilde{G}}, \dots, g_{|\tilde{G}|}^{\tilde{G}}, g_1^{\hat{G}}, g_2^{\hat{G}}, \dots, g_{|\hat{G}|}^{\hat{G}}\}$, where $\omega_1(.)$ retrieves all similar findings of $t$ but it does so in a particular manner, where the ranking is such that first all elements of $\tilde{G}$ are retrieved and then all elements of $\hat{G}$. Considering that our main metrics of interest, MAP and MRR, are rank-sensitive, the perfect IR system will get non-perfect scores due to the preference of elements from $\tilde{G}$ over $\hat{G}$.
Our validation approach is then applied to construct $M=1000$ down-sampled versions ($\bar{d}_i$) of $db$, with $m=100$. The IR system is applied to each of these $\bar{d}_i$, which may or may not contain some samples belonging to $\tilde{G}$, given the finite size of $db$. The average scores yielded by the IR system with respect to MAP and MRR are then computed.
This exercise is repeated $mc=10,000$ times for different ranges of $|\tilde{G}|=\{5,10,15,20\}$ while fixing $|\hat{G}|=3$\footnote{The chosen values are based on a rough investigation of the findings database $F$ and expert judgement.}. The averages across the runs serve as the theoretical upper bounds of the scores that can be achieved.
This strategy allows us to validate our models on a much larger scale than manual labelling of the entire database would allow, while still providing an accurate estimate of the model's performance. It effectively reduces the possibility of misjudging an optimal IR system due to unidentified relevant findings, thus making the performance evaluation more fair and reliable.

\begin{figure}[hp]
 \centering
 \includegraphics[width=.99\textwidth]{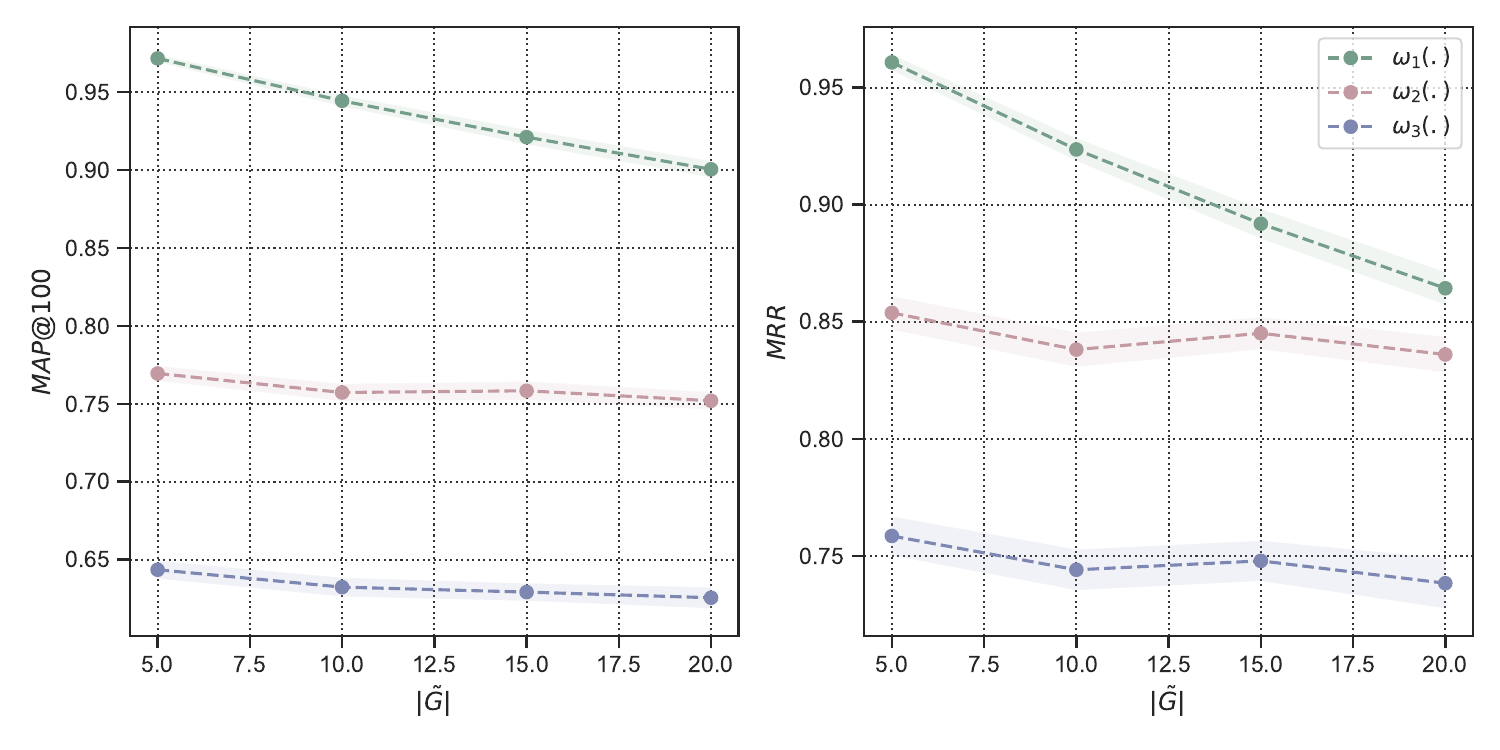}
 \caption{Simulation exercise of a perfect IR system, $\omega_1(.)$. $\omega_2(.)$ is a biased system that samples from $G$ with a higher probability than from $D\backslash G$. $\omega_3(.)$ is a similar system, albeit with a lower likelihood of sampling from $G$ compared to $\omega_2(.)$.}
 \label{fig:val}
\end{figure}

The outcomes of the simulation exercise are depicted in Figure~\ref{fig:val}. The left panel illustrates the theoretical upper bounds of MAP scores across various values for $|\Tilde{G}|$, while the right panel mirrors this depiction for MRR. 

Additionally, we consider two other systems: $\omega_2(.)$ and $\omega_3(.)$. $\omega_2(.)$ is a biased system that samples from $G$ with a higher probability than from $D\backslash G$. $\omega_3(.)$ is a similar system, albeit with a lower likelihood of sampling from $G$ compared to $\omega_2(.)$. Under the premise of a valid evaluation methodology, the performance hierarchy should ideally be: $\omega_1(.)$ outperforming $\omega_2(.)$, which in turn outperforms $\omega_3(.)$. The results indeed confirm this expected order, thereby validating our evaluation approach.

The performance of the perfect system, evident from both MAP and MRR metrics, exhibits a monotonic decrease in response to an increase in $|\Tilde{G}|$. As $|\Tilde{G}|$ increases, while $|db|$ is fixed, the probability of drawing samples from $|\Tilde{G}|$ when constructing a down sampled version $db_i$ of the database increases, leading the IR system to be wrongly penalized.

The theoretical upper limit for MAP scores ranges between 97\% (for $|\Tilde{G}|=5$) and 90\% (for $|\Tilde{G}|=20$). Similarly, the theoretical upper bound for MRR ranges from 97\% (for $|\Tilde{G}|=5$) to 87\% (for $|\Tilde{G}|=20$). These values are instrumental in the evaluation of the performance of our models on the actual database $F$.

\section{Results \& Discussion} \label{sec:result}

Table~\ref{tab:performance_metrics} presents the results of the evaluation methodology applied to different variations of the IR system, depending on the underlying embedding scheme applied. For every variant of the IR system (enumerated in rows), we calculate the \textit{Mean Average Precision} at 100 (\textit{MAP@100}), the \textit{Mean Reciprocal Rank} (\textit{MRR}), and the aggregate mean score (\textit{avg score}). From the table, it is evident that the performance of IR systems significantly varies, indicating that the choice of embedding can markedly affect the efficiency and accuracy of information retrieval tasks.

\textbf{Lexical}

Starting with baseline models, the \textit{Random} model exhibits the lowest performance across all metrics (MAP=0.07 and MRR=0.11), serving as a control to underscore the effectiveness of more sophisticated approaches. Among traditional models, TFIDF and variations of BM25 showed substantial improvements. Notably, BM25L+ outperformed other variations with a MAP of 0.87 and an MRR of 0.70, demonstrating the effectiveness of optimizations on the basic BM25 model for retrieving relevant findings. This model incorporated length normalization, pruned token frequency matrix based on the document frequency of the tokens, as well as bi- and trigrams, to better handle variations in document length and term frequency. 

\textbf{Word-level Embeddings}

Moving on to semantic approaches, the word embedding models GloVe (MAP=0.39, MAP=0.66) and FinGloVe (MAP=0.37, MRR=0.64) demonstrate less satisfactory results. The sub-optimal performance of GloVe can be largely attributed to insufficient training on prudential data, leaving it unable to embed specialized terms like "\textit{probability of default}" effectively \cite{pennington2014glove}. This deficiency illustrates a critical gap in vocabulary that hinders the model's effectiveness in specialized domains. Despite being trained on financial data, FinGloVe performs even worse, echoing concerns in previous studies about the detrimental effect of limited training data on the effectiveness of word embedding models \cite{mikolov2013efficient}. The limited scope of its training data, which covered a narrow range of prudential documents relative to the pretrained GloVe model, likely contributed to its underperformance by not providing a sufficiently diverse linguistic context.

The BERT-based systems underperform as a group. BERT (MAP=0.34, MRR=0.60) and FinBERT (MAP=0.33, MRR=0.61) register the lowest scores in MRR and MAP metrics, respectively. These models' limitations could be attributed to the complex and domain-specific language of the prudential documents, which these models might struggle to comprehend \cite{devlin2018bert}. However, LegalBERT (MAP=0.39, MRR=0.68)), which is trained on legal domain data, shows a slight improvement, suggesting that further training on domain-related data might yield more positive results \cite{chalkidis2020legalbert}.

\textbf{Document-level Embeddings}

In contrast, the sentence transformers show more promise. The base sentence transformer, \textit{SentTRF}, outperforms the BERT-like models (MAP=0.47, MRR=0.69), implying that pretrained sentence transformers might have learned to create meaningful embeddings capturing the global context of findings to some extent \cite{reimers2019sentencebert}. Nonetheless, its performance is still considerably lower than the fully configured \textit{BM25L+} model.

However, when trained on the CRR and historical findings $F$ (based on \textit{TSDAE}), the sentence transformer's performance improves significantly (MAP=0.72, MRR=0.88), reinforcing the importance of domain-specific training in generating high-quality embeddings. These embeddings surpass \textit{BM25L+} significantly in terms of MAP and slightly in MRR.

The hybrid model, \textit{Hybrid}, combining the strengths of lexical matching (\textit{BM25L+}) and semantic understanding (\textit{SentTRF+TSDAE}) delivers the most encouraging results. The hybrid model achieves a MAP score of 0.74 and MRR score of 0.89, outperforming all other models. This suggests that combining multiple retrieval strategies and model architectures can lead to superior performance in IR systems. More specific, combining lexical and semantic techniques in prudential data can enhance IR system performance, aligning with research supporting integrated methods \cite{9321861}.

\begin{table}[htbp]
  \centering
  \caption{Performance Metrics across different IR-system variations}
  \label{tab:performance_metrics}
  \begin{tabular}{lccc}
    \toprule
    \textbf{Model} & \textbf{MAP@100} & \textbf{MRR@100} & \textbf{avg score} \\
    \midrule
    Random & 0.0708 & 0.1124 & 0.0916 \\
    \midrule
    TFIDF & 0.6296 & 0.8218 & 0.7257 \\
    BM25 & 0.6458 & 0.8487 & 0.7473 \\
    BM25Plus & 0.6440 & 0.8539 & 0.7489 \\
    BM25L & 0.6466 & 0.8543 & 0.7504 \\
    BM25L+ & 0.6957 & 0.8722 & 0.7840 \\
    \midrule
    GloVe & 0.3938 & 0.6570 & 0.5254 \\
    FinGloVe & 0.3726 & 0.6413 & 0.5069 \\
    \midrule
    BERT & 0.3410 & 0.5959 & 0.4685 \\
    FinBERT & 0.3283 & 0.6063 & 0.4673 \\
    Legal-BERT & 0.3874 & 0.6838 & 0.5356 \\
    \midrule
    SentTRF & 0.4678 & 0.6904 & 0.5791 \\
    SentTRF+TSDAE & 0.7173 & 0.8750 & 0.7961 \\
    Hybrid & \textbf{0.7425} & \textbf{0.8925} & \textbf{0.8175} \\
    \midrule
    Perfect model $\omega_1(.)$ & 0.9383 & 0.9455 & 0.9419 \\
    \bottomrule
  \end{tabular}
\end{table}

\

Table~\ref{tab:performance_metrics_crr} presents the results of incorporating the fuzzy matching component into the different variations of the IR system. The inclusion of fuzzy matching, particularly evident in the \textit{Random} model (MAP=0.57, MRR=0.63), demonstrates the substantial improvement achieved by leveraging this approach. By employing fuzzy matching based on CRR articles, the search space is refined, providing contextual relevance and aiding both lexical and semantic models. Lexical models such as TFIDF and BM25 significantly benefit from operating within this refined space, evidenced by their enhanced performance metrics: TFIDF attains a MAP of 0.79 and an MRR of 0.88, while BM25 achieves a MAP of 0.80 and an MRR of 0.91. The fine-tuned BM25L+ model shows the best performance among lexical models, with a MAP of 0.82 and an MRR of 0.93, demonstrating the effectiveness of fuzzy matching in enhancing the precision of these models. 

For semantic models, word embeddings and BERT-based systems also leverage the narrowed semantic understanding task to enhance their performance. For instance, GloVe and FinGloVe show improved results with MAPs of 0.71 and 0.70, and MRRs of 0.84 and 0.87, respectively. However, FinGloVe, despite its specialized training, slightly lags behind GloVe in MAP. BERT-based models like LegalBERT benefit significantly from fuzzy matching, with a MAP of 0.72 and an MRR of 0.89, outperforming both BERT and FinBERT.

The domain-adapted Sentence Transformer, \textit{SentTRF+TSDAE}, substantially benefits from fuzzy matching, achieving a MAP of 0.82 and an MRR of 0.92. The \textit{Hybrid} model, which combines the strengths of lexical and semantic models optimized with fuzzy matching, records the highest performance metrics of all variations: a MAP of 0.83 and an MRR of 0.93.

\begin{figure}[ht!]
    \centering
    \includegraphics[width=0.99\textwidth]{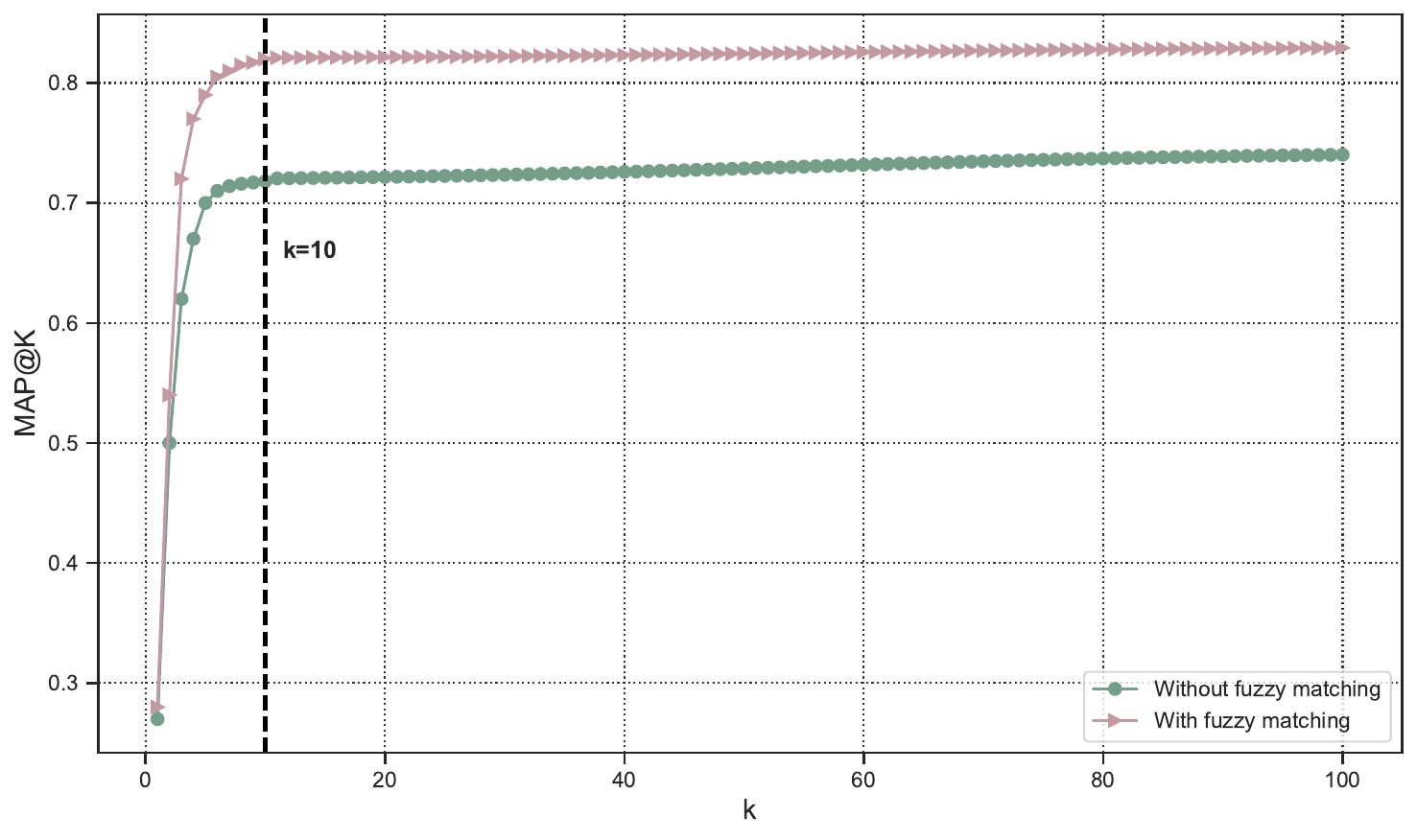}
    \caption{$MAP$ across different $k$ values for the \textit{Hybrid} model}
    \label{fig:fuzzy_vs_nofuzzy}
\end{figure}

The insights gained from Figure~\ref{fig:fuzzy_vs_nofuzzy}, which charts the MAP scores against varying $k$ values for the \textit{Hybrid} model, further substantiate these findings. The application of fuzzy CRR matching uplifts the performance of the model almost uniformly.  Interestingly, only a marginal increase in MAP scores is observed beyond $k=10$, suggesting that retrieving the top 10 similar findings should suffice for the majority of usecases.

\begin{table}[htbp]
  \centering
  \caption{Performance Metrics across different IR-system variations (with CRR matching)}
  \label{tab:performance_metrics_crr}
  \begin{tabular}{lccc}
    \toprule
    \textbf{Model} & \textbf{MAP@100} & \textbf{MRR@100} & \textbf{avg score} \\
    \midrule
    Random & 0.5688 & 0.6317 & 0.6003 \\
    \midrule
    TFIDF & 0.7868 & 0.8836 & 0.8352 \\
    BM25 & 0.8014 & 0.9109 & 0.8561 \\
    BM25Plus & 0.8035 & 0.9178 & 0.8607 \\
    BM25L & 0.8031 & 0.9157 & 0.8594 \\
    BM25L+ & 0.8218 & 0.9266 & 0.8742 \\
    \midrule
    GloVe & 0.7145 & 0.8426 & 0.7786 \\
    FinGloVe & 0.7028 & 0.8667 & 0.7848 \\
    \midrule
    BERT & 0.6982 & 0.8452 & 0.7717 \\
    FinBERT & 0.6768 & 0.8015 & 0.7392 \\
    LegalBERT & 0.7222 & 0.8857 & 0.8040 \\
    \midrule
    SentTRF & 0.7255 & 0.8304 & 0.7780 \\
    SentTRF+TSDAE & 0.8199 & 0.9202 & 0.8700 \\
    Hybrid & \textbf{0.8301} & \textbf{0.9274} & \textbf{0.8788} \\
    \midrule
    Perfect model $\omega_1(.)$ & 0.9383 & 0.9455 & 0.9419 \\
    \bottomrule
  \end{tabular}
\end{table}


As a last evaluation method, we conduct the following performance evaluation, to enact a realistic retrieval scenario. Given $7$\footnote{due to time-constraints, the number of input findings is kept small.} handpicked findings from $F$, we ask the top performing variations, to retrieve the top $k=10$ similar findings. We then ask experts in the field to manually evaluate the retrieved findings for each of the models. Afterwards we compute the $MRR$ of the models. The results are depicted in table~\ref{table:man_ex}. \textit{Hybrid} is the top performer with an $MRR$ score of 82\%, indicating that on average a similar finding is retrieved within the first 3 ranked findings.

\begin{table}[ht]
\centering
\caption{$MRR$ performance of manual evaluation exercise}
\label{table:man_ex}
\begin{tabular}{lc}
\toprule
 & MRR \\
\midrule
BM25L+ & 0.75 \\
SentTRF+TSDAE & 0.76\\
Hybrid & \textbf{0.82} \\
\bottomrule
\end{tabular}
\end{table}

\section{Limitations \& Future Work}
While the proposed IR system offers substantial benefits, it is important to recognize its limitations. First, the evolving regulatory landscape poses challenges, as changing regulations can make it difficult for supervisors to locate historically similar findings and measures. Addressing this requires more than minor adjustments; it necessitates an auxiliary system that adapts to regulatory changes, with regular updates and continuous learning mechanisms. Second, the system retrieves similar past findings and associated measures, but there is potential to enhance it by approximating the mapping \(\theta:F \rightarrow P_{\geq 1}(M)\). A well-defined approximation, $\hat{\theta}$, could suggest new measures based on inputted findings. Rewriting the problem as a Neural Machine Translation (NMT) task, using a generative Large Language Model to "translate" findings into measures, is a promising approach. Conditioned on past measures, the NMT model would ensure consistency with historical data, offering dynamic and contextually appropriate responses. However, this introduces complexities and potential inaccuracies that must be thoroughly validated.

\section{Conclusion}\label{conclusion}

In this paper, we introduce a novel Information Retrieval (IR) system that leverages Large Language Models (LLMs) to support supervisors in drafting consistent and effective measures within the Single Supervisory Mechanism (SSM) framework. By integrating lexical, semantic, and fuzzy set matching techniques, the system enhances the ability of supervisors to draft measures grounded in a robust analytical comparison with historical data.

Our results demonstrate that the IR system significantly improves the efficiency and quality of supervisory decision-making by providing supervisors with relevant historical contexts and parallels, thereby reducing time spent on manual searches and increasing the accuracy of measure formulation. The use of Monte Carlo validation methods has further confirmed the system's effectiveness, even in scenarios characterized by partially labeled data, showcasing its robustness and adaptability.



\bibliographystyle{elsarticle-harv} 
\bibliography{sn-bibliography}





\end{document}